\documentclass{paper}
\usepackage{epsfig, graphicx}
\usepackage[english]{babel}

\begin{document}
\title{Multilayer of Phospholipid Membranes on a Hydrosol Substrate}
\author{Aleksey M. Tikhonov\/\thanks{tikhonov@kapitza.ras.ru}}
\maketitle
\leftline{\it Kapitza Institute for Physical Problems, Russian Academy of Sciences,}
\leftline{\it ul. Kosygina 2, Moscow, 119334, Russia}
\rightline{\today}

\abstract{The molecular structure of a multilayer of 1,2-distearoyl-sn-glycero-3-phosphocholine (DSPC) adsorbed on
the surface of the hydrosol of silica nanoparticles has been studied by the synchrotron radiation scattering
method. According to the reflectometry data, the multilayer is formed by planar phospholipid bilayers with a
thickness of $69 \pm 1$ $\rm \AA$ and its total thickness is about 400 \AA. Grazing-incidence diffraction indicates that the bilayers are in the crystal state with an area of $41.6 \pm 0.7$ $\rm \AA^2$ per molecule.}
\vspace{0.25in}

\large
Phosphatidylcholine molecules in an aqueous medium form microscopic aggregates, vesicles and
liposomes, whose shell is formed by molecular bilayers and is often considered as a model of a biological
membrane [1]. Detailed information on the structure of a lipid bilayer is important for the solution of a large
number of biophysical problems. The characteristic spontaneous curvature radius of a phospholipid membrane in the aqueous medium is smaller than 10 $\mu \rm m$; for this reason, the membrane samples for structural
investigations are prepared on planar solid substrates made of crystalline silicon, glass, quartz, or polymer
[2 -- 7]. In this paper, a method for obtaining macroscopically planar lipid membranes on a liquid
substrate made of the aqueous solution of amorphous silica particles is reported.

The film samples of 1,2-distearoyl-sn-glycero-3-phosphocholine (DSPC, see Fig. 1)
were prepared and studied in Teflon dish with a diameter of
about 100 mm. Using a 25 $\mu \rm l$ syringe (Hamilton), a 5 -
to 10 - $\mu \rm l$ drop of the solution of phospholipid in chloroform ($\sim 0.03$ mole/l)
is deposited on the surface of the liquid substrate; the amount of the substance in the
drop is enough to form a multilayer consisting of more
than ten lipid monolayers. In this case, the spreading
of the drop on the surface is accompanied by a
decrease in the surface tension $\gamma$ of the air - hydrosol
interface from 74 to 35 - 40 mN/m, which was
detected by the Wilhelmy method using an NIMA PS-2 surface pressure gauge.
Then, the equilibrium of the sample was reached inside an air-tight single-stage thermostat at T =
298 K for about 12 h.

For comparison, an adsorbed DSPC monolayer
was similarly prepared on the aqueous substrate ($\rm pH \approx 12$), which is a NaOH solution
(99.95 \% in the metal content, Sigma-Aldrich) in deionized water (Barnstead UV).
In this case, $\gamma \sim 50$ mN/m.

\begin{figure}
\hspace{0.5in}
\epsfig{file=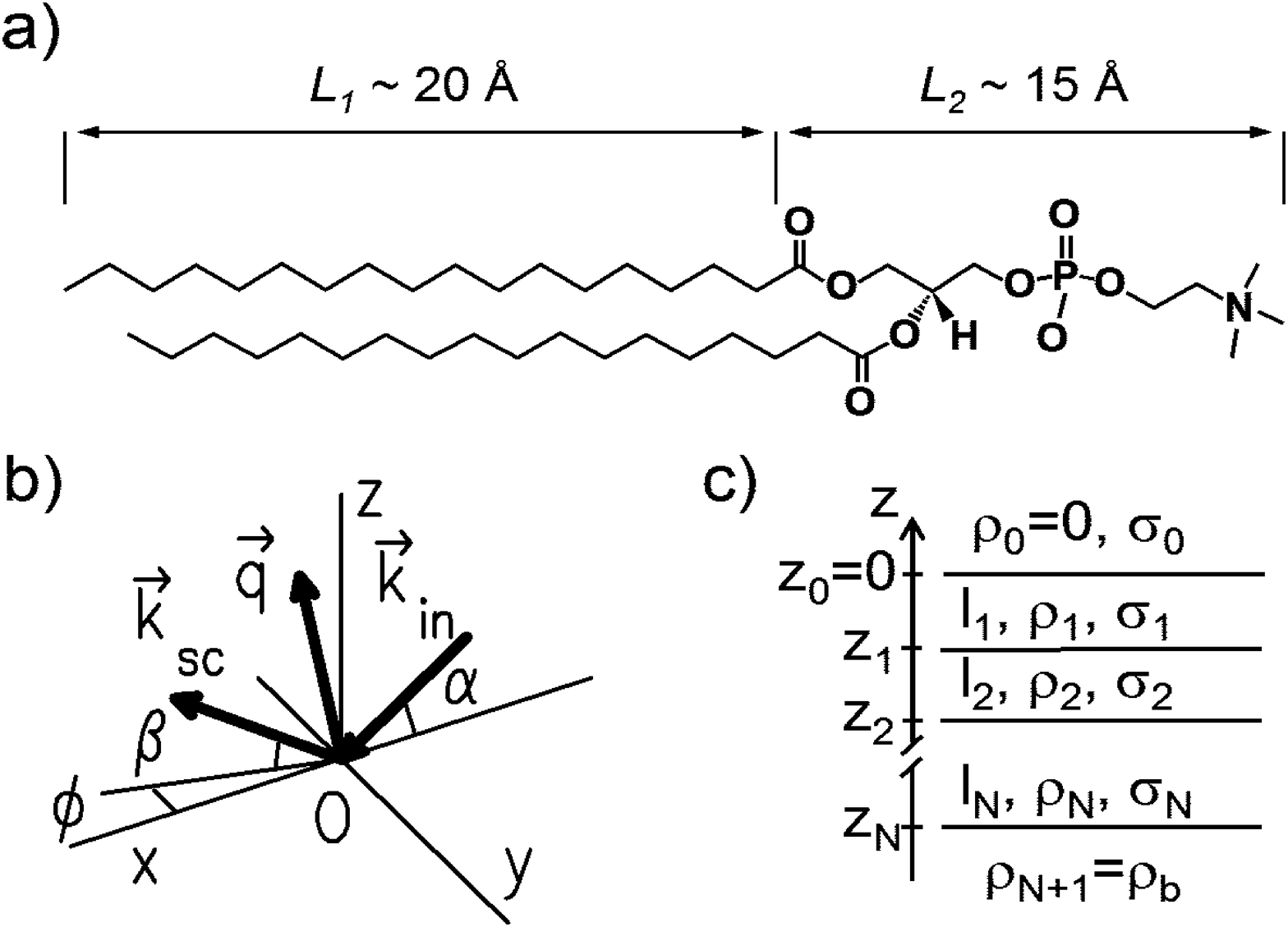, width=0.75\textwidth}

Figure 1. (a) Structure of 1,2-distearoyl-sn-glycero-3-phosphocholine (DSPC);
(b) the kinematics of scattering on the surface of a liquid,
where ${\bf k}_{in}$ is the wave vector of the
incident wave, ${\bf k}_{sc}$ is the wave vector of the wave scattered
in the direction of the observation point, and ${\bf q} = {\bf k}_{in} -{\bf k}_{sc}$
is the scattering vector; and (c) the parameterization of the
interface electron density $\rho(z)$. The adsorbed layer is
divided into $N$ layers (slabs) with thickness $l_{j}$ and electron
density $\rho_j$ (slab model); $\sigma_j = \sigma_0$ is the standard deviation of
the position of the $j$th interface from the nominal value $z_j$.
The layer adjacent to the substrate is formed by polar groups of glycero-3-phosphocholine.
The surface and bulk densities are $\rho_N = \rho_w$ and $\rho_{N + 1} = \rho_b$, respectively.
The contribution from the transition region to the reflectivity at
$q_z > 0.1 \rm \AA^{–1}$ is neglected ($\sigma_N >> \sigma_0$) [16].

\end{figure}

A powder of synthetic 1,2-distearoyl-sn-glycero-3-phosphocholine was purchased from Avanti
Polar Lipids and chloroform ($\sim 99.8\%$) was purchased from Sigma-Aldrich.
Monodisperse hydrosols of amorphous silica particles with a diameter of about 220 \AA were
delivered by Grace Davison. These solutions (${\rm pH} \approx 9$) with a density of
($1.30 \pm 0.01$) g/cm$^3$ (Ludox TM-40, $\sim 40$ wt \% of SiO$_2$) and ($1.40 \pm 0.01$) g/cm$^3$
(Ludox TM-50, $\sim 40$ wt \% of SiO$_2$) were stabilized by sodium hydroxide ($\sim 0.2$ mole/l).

The X-ray diffraction data were obtained on the X19C station at the National Synchrotron
Light Source (NSLS) [8].
The transverse and in-plane structures of the lipid films were investigated by the methods
of reflectometry and grazing-incidence diffraction of a focused beam ($\sim 10^{11}$ photons/s)
with a photon energy of 15 keV ($\lambda = 0.825 \pm 0.002$ ${\rm \AA}$), respectively.

\begin{figure}
\hspace{0.5in}
\epsfig{file=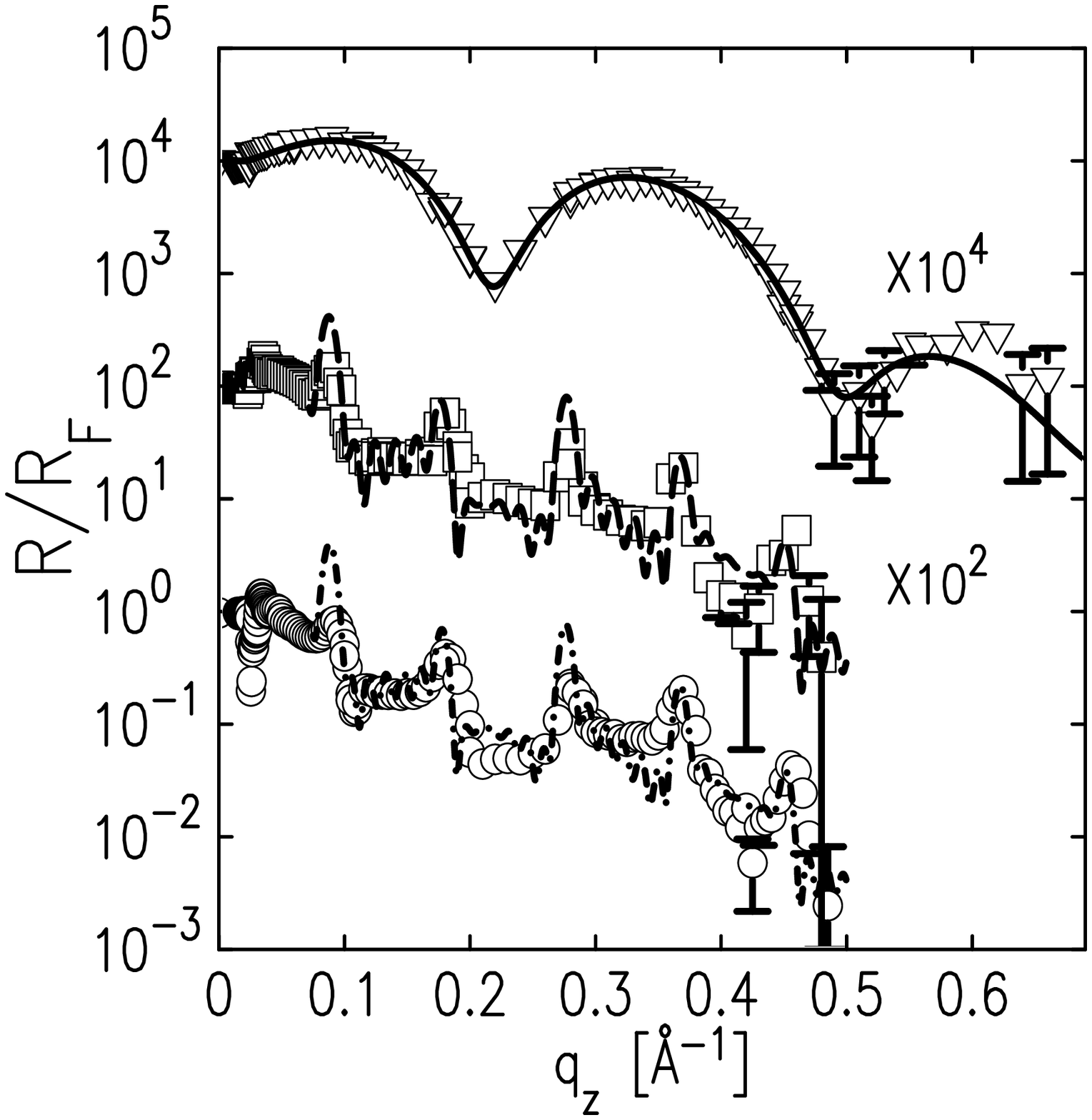, width=0.75\textwidth}

Figure 2. Ratio $R(q_z)/R_F(q_z)$ for (triangles) the DSPC monolayer on the surface of water;
the solid line is the two-layer model of the monolayer; and (squares and circles) the lipid
multilayer on the surface of hydrosol with a silica concentration of 40 and 50 wt \%, respectively;
the dashed line is the model of six bilayers on the substrate with
$\rho_b=1.21 \rho_w$ and $\sigma_0 =4.4$ {\AA}; and the dash–dotted line is the model of
seven bilayers on the substrate with $\rho_b=1.3 \rho_w$ è $\sigma_0 =4.2$ {\AA}.
\end{figure}

The kinematics of scattering on the surface of the liquid is naturally described in the
 system of coordinates with the $xy$ plane coinciding with the surface and
the $z$ axis normal to the surface and opposite to the gravitational force (see Fig. 1b).
The triangles in Fig. 2 present the coefficient of reflection $R$ as a function of
$q_z=(4\pi/\lambda)\sin\alpha$ for the DSPC monolayer adsorbed on
the surface of water. The squares and circles in Fig. 2 present the coefficient of reflection of the phosphocholine films adsorbed on the surface of hydrosol with silica concentrations of 40 and 50 \%, respectively. For
a more convenient representation of the $R(q_z)$ dependence, it is normalized to the Fresnel function
$R_F(q_z)\approx$ $(q_z-[q_z^2-q_c^2]^{1/2})^2/(q_z+[q_z^2-q_c^2]^{1/2})^2$, where $q_c=(4\pi/\lambda)\sin\alpha_c$.
At the glancing angles smaller than $\alpha_c\approx\lambda\sqrt{r_e\rho_b/\pi}$ (where $r_e = 2.814\cdot10^{-5}$ {\AA} is the classical electron radius and $\rho_b$ is the average electron density in the substrate), the incident ray undergoes total reflection; i.e., $R \approx 1$. The electron density of water $\rho_w\approx 0.333$  {\it e$^-$/}{\AA}$^3$ (where {\it e$^-$} is the elementary charge); the electron densities of Ludox TM-40 and Ludox TM-50 silica hydrosols are $\rho_b \approx 1.2 \rho_w$ (Ludox TM-40) and  $\rho_b \approx 1.3 \rho_w$(Ludox TM-50), respectively.

The period of oscillations in $R(q_z)$ for the aqueous substrate is about twice as long as that for the hydrosol
substrate. Moreover, the $R(q_z)/R_F(q_z)$ dependence of the surface of hydrosol contains a set of sharp peaks,
whereas the structure factor of the monolayer on the surface of water is characterized by wide oscillations.

The quantity $R(q_z)$ contains information on the transverse profile of the electron density $\rho (z)$ averaged
over a large area ($\sim 0.5$ cm$^2$). The density $\rho(z)$ is parameterized in the standard (slab) model of the multilayer (see Fig. 1b). The model profile is constructed on the basis of the error function under the assumption that $\sigma_j$ is the standard deviation of the position of the $j$-th interface of the multilayer from the nominal value $z_j$ [9].

The lower limit of the parameters $\sigma_j$ is determined by the capillary width,
$\sigma_{cw}^2 = ( k_BT/2\pi\gamma ) \ln(Q_{max}/Q_{min})$ (where $k_B$ is the Boltzmann constant),
which is specified by the short-wavelength limit in the spectrum of capillary waves
$Q_{max} = 2\pi/a$ (where $a\approx 5$ {\AA} is the inter molecular distance, and $Q_{min}=q_z^{max}\Delta\beta $ ($2\Delta\beta$$\approx 6\cdot10^{-4}$ rad is the angular resolution of the detector and $q_z^{max} \approx 0.5$ {\AA}$^{-1}$) [10-13]. Under the assumption that $\sigma_j=\sigma_0$ for all $j$ values,
the structure factor in the first Born approximation has the form [14]
\begin{equation}
\frac{R(q_z)}{R_F(q_z)} \approx  \left| \frac{1}{\rho_b}\sum_{j=0}^{M}{(\rho_{j+1}-\rho_j) e^{-iq_zz_j}} \right|^2 e^{-\sigma_0^2q_z^2}.
\end{equation}

Analysis of the data presented by triangles in Fig. 2
indicates that DSPC molecules are adsorbed as a monolayer on the surface of water;
the coefficient of reflection of this monolayer is well described by the
two-layer model with five fitting parameters.
The variation of the parameters in the monolayer model is in
agreement with the structure of the DSPC molecule (see Fig. 1a).
The first layer of the thickness $L_1 \sim 18$ $\rm \AA$  is formed by
hydrocarbon chains and has the density $\rho_1=1.01 \rho_w$, which corresponds to the densest packing
of saturated hydrocarbons in the $\gamma-$phase [1]. The second layer of the monolayer,
which is in direct contact with water, is formed by polar hydrophilic groups of
glycerophosphocholine and has the density $\rho_2=1.38 \rho_w$ and the thickness $L_2 \sim 9$ {\AA}.
The calculated value $\sigma_{cw}=3.4 \pm 0.2$ {\AA} coincides within the errors with the fitted
value $\sigma_{0}=3.6 \pm 0.2$ {\AA}. The solid line in Fig. 2 is the model structure factor with
the parameters presented in the table. The $\rho(z)$ profile of the two-layer model of
the monolayer is shown in Fig. 4a (see below) and is in agreement with the previous results [15].

The phospholipid film adsorbed at the hydrosol's surface is a multilayer. At $q_z > 0.1$ \AA$^{-1}$,
scattering on the inhomogeneities of the electron density, which are
attributed to the transverse distribution of 22-nm silica
particles, makes a small contribution to the reflected
power [16]; i.e., $R(q_z)$ at these angles is determined
only by the structure of the lipid layer. The periodicity
of peaks $\Delta Q_z \sim 0.09$ \AA$^{–1}$ in $R(q_z)$ in Fig. 2 corresponds
to the layered structure along the $z$ axis with the period
$L ~ 2\pi/\Delta Q_z \sim 70$ \AA, which is the thickness of the double layer of DSPC molecules.
A relatively small width of the peaks, $\delta q_z$ $\sim 1.5 \cdot10^{-2}$ {\AA}$^{-1}$  in $R(q_z)$
is due to the interference of reflected rays in the layer about
$\sim$ 400 {\AA} in thickness, which can contain $\Delta Q_z/\delta q_z \sim 6 - 7$ lipid
bilayers.

\begin{figure}
\hspace{0.5in}
\epsfig{file=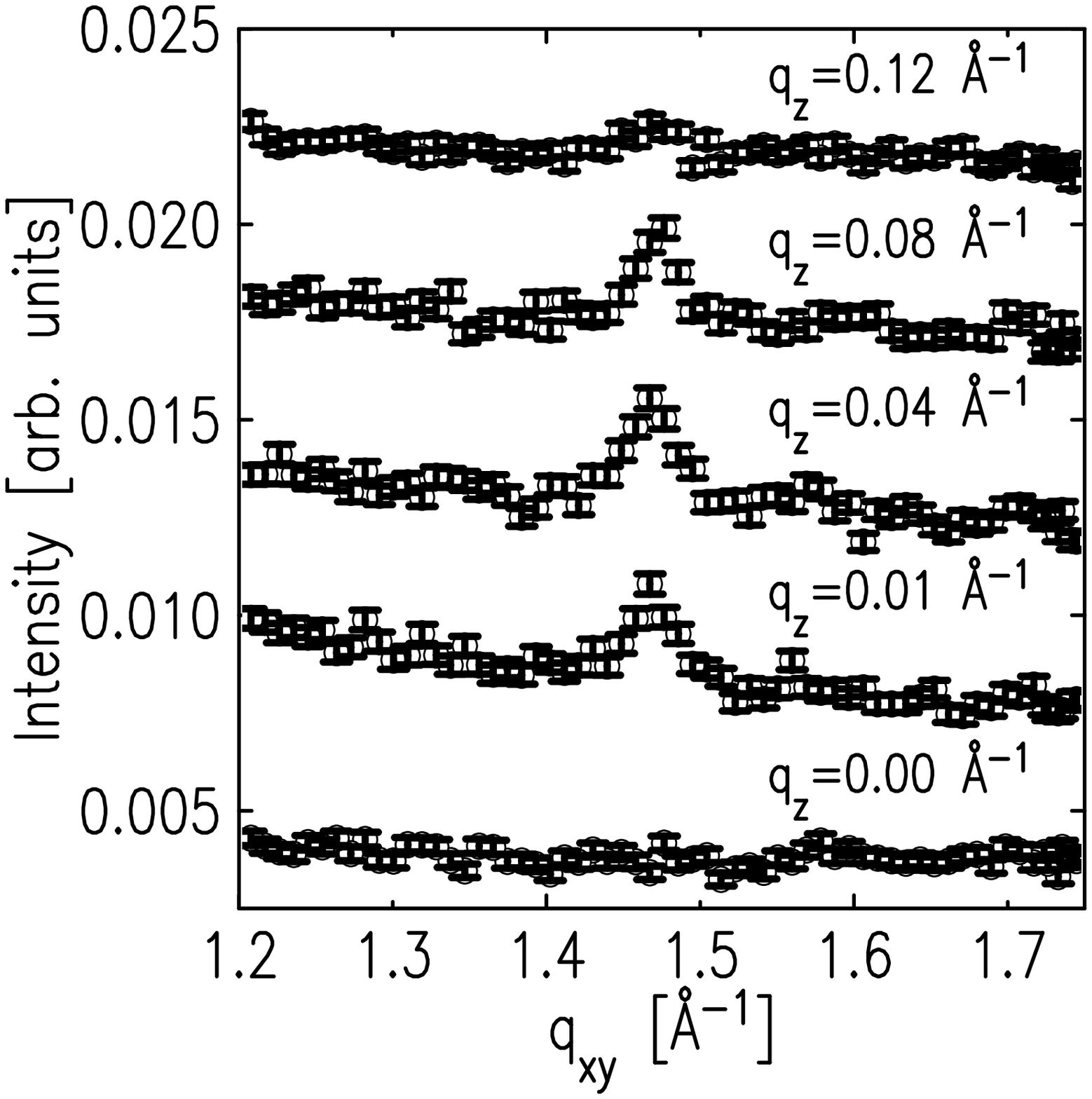, width=0.75\textwidth}

Figure 3. Intensity of grazing-incidence diffraction from the lipid multilayer at
the air - hydrosol interface. The Bragg peak structure along the $z$ axis
was measured at the glancing angle
$\alpha = 0.07$ deg. using the vertical position-sensitive detector
(Ordela). The horizontal angular resolution of the detector,
which determines the width of the diffraction peak, is
about 0.003 rad or $\Delta q_{xy} \approx 0.02$ \AA$^{-1}$.
The angular resolution of the detector in the vertical plane is about 0.002 rad or
$\Delta q_z \approx 0.02$ \AA$^{-1}$.
\end{figure}

The presence of only one peak in the grazing-incidence diffraction (see Fig. 3) indicates the highly symmetric
hexagonal packing of hydrocarbon tails in the multilayer.
It's position at $q_{xy} = 1.48 \pm 0.01$ \AA$^{-1}$
($q_{xy}=(q_x^2+q_y^2)^{1/2}$) corresponds to the triangular lattice of the
hydrocarbon chains with the period $a=4.91\pm 0.04$ \AA.
The area per chain, $S=20.8 \pm 0.5$ {\AA}$^{2}$, within the measurement
error coincides with the value in the hexagonal bulk phase $P_{\beta '}$
of phosphocholine crystals [1].

\small
\vspace{5mm}
Table. Model parameters of the electron-density profile (see Fig. 4)
\vspace{2mm}

\hspace{-8mm}
\begin{tabular}{|c|c|c|c|c|c|}
\hline
&&&&&\\
Substrate &$L_1$({\AA})&$L_2$({\AA})& $\rho_1 / \rho_w$ & $\rho_2 / \rho_w$ & $\sigma_0$({\AA})\\
&&&&& \\
\hline
&&&&& \\
Water & $17.5^{+3/-0.5}$ & $9 ^{+1/-5}$& $1.005^{+0.025/-0.006}$ & $1.38^{+0.3/-0.05}$ & $3.6 ^{+0.2/-0.1}$\\
&&&&& \\
Hydrosol &$20 ^{\pm 1}$ & $15 ^{\pm 1 }$& $0.97 ^{\pm 0.05}$& $0.80 ^{\pm 0.02}$ & $4.2 ^{\pm 0.2}$\\
&&&&& \\
\hline
\end{tabular}
\vspace{2mm}

Note: $L_1$ is the thickness of the layer of hydrocarbon tails with the density $\rho_1$, $L_2$ is the thickness of the layer of polar groups with the density $\rho_2$,
 and $\sigma_0$ is the width of the interfaces; the densities are given in units of the density of water $\rho_w=0.333$  {\it e-/}{\AA}$^3$. The errors of the determination of the parameters were obtained using the standard $\chi^2$ criterion at the confidence level of 0.9
\large
\vspace{5mm}

The positions and intensities of the peaks in $R(q_z)$ can be satisfactorily
described by the electron density profile specified by only five parameters,
which can be easily determined using the X-ray diffraction data for
phosphocholine crystals. The volume occupied by the
-CH$_2$- group in the hexagonal phase is $V_0 \sim 25 - 26$ {\AA}$^3$
[1, 17]. Thus, the thickness of the layer of hydrocarbon
tails CH$_3$-(CH$_2$)$_{16}$- of DSPC is $L_1\sim 17V_0/S$ $\approx 20$ {\AA} and
the electron density of this layer is $\rho_1 \approx \rho_w$ (137 electrons in the volume
$L_1 S$). Since $L=2(L_1+L_2)$, the
thickness of the layer formed by polar groups is $L_2=L/2-L_1$ $\sim 15$ {\AA} and
its density is determined by 164 electrons of the glycerophosphocholine group $\rho_2 = 164e^-/(2SL_2)$ $\approx 0.8 \rho_w$ (the C$_{44}$H$_{88}$NO$_8$P molecule contains 438 electrons).

The dashed and dash–dotted lines in Fig. 2 correspond to the structure factors given by Eq. (1) for the
model of the multilayer formed by the six and seven bilayers, respectively (see Fig. 4b, table). In these
experiments, a decrease in $R(q_z)$ at large $q_z$ values is well described by the calculated value
$\sigma_0=4.2 \pm 0.2$ {\AA}. A better quantitative agreement of the model with the
experimental data at $q_z > 0.05$  {\AA}$^{-1}$ can apparently be achieved by taking into
account the presence of topological defects of the occupation of the multilayer at
the interface with air. However, in this case, the certainty of the description will require exact knowledge
of the morphology of the surface.

Thus, DSPC molecules are adsorbed as a monolayer on the surface of water,
whereas they form a multilayer of membranes on the surface of hydrosol.
The estimate shows that the areas per lipid molecules both
in the monolayer on the surface of water and in the
bilayer are approximately the same, 42 \AA$^2$.
However, according to the parameters of the models presented
in the table, the layer of polar groups in the monolayer
is $\sim 50\%$ denser and $\sim 50\%$  thinner than that in the
bilayer. This indicates that the configurations of glycerophosphocholine groups are
different in these layers. This is possibly due to the hydration of the
polar group of the DSPC molecule on the surface of
water, whereas lipid molecules in the multilayer apparently remain dry [18, 19].

\begin{figure}
\hspace{0.5in}
\epsfig{file=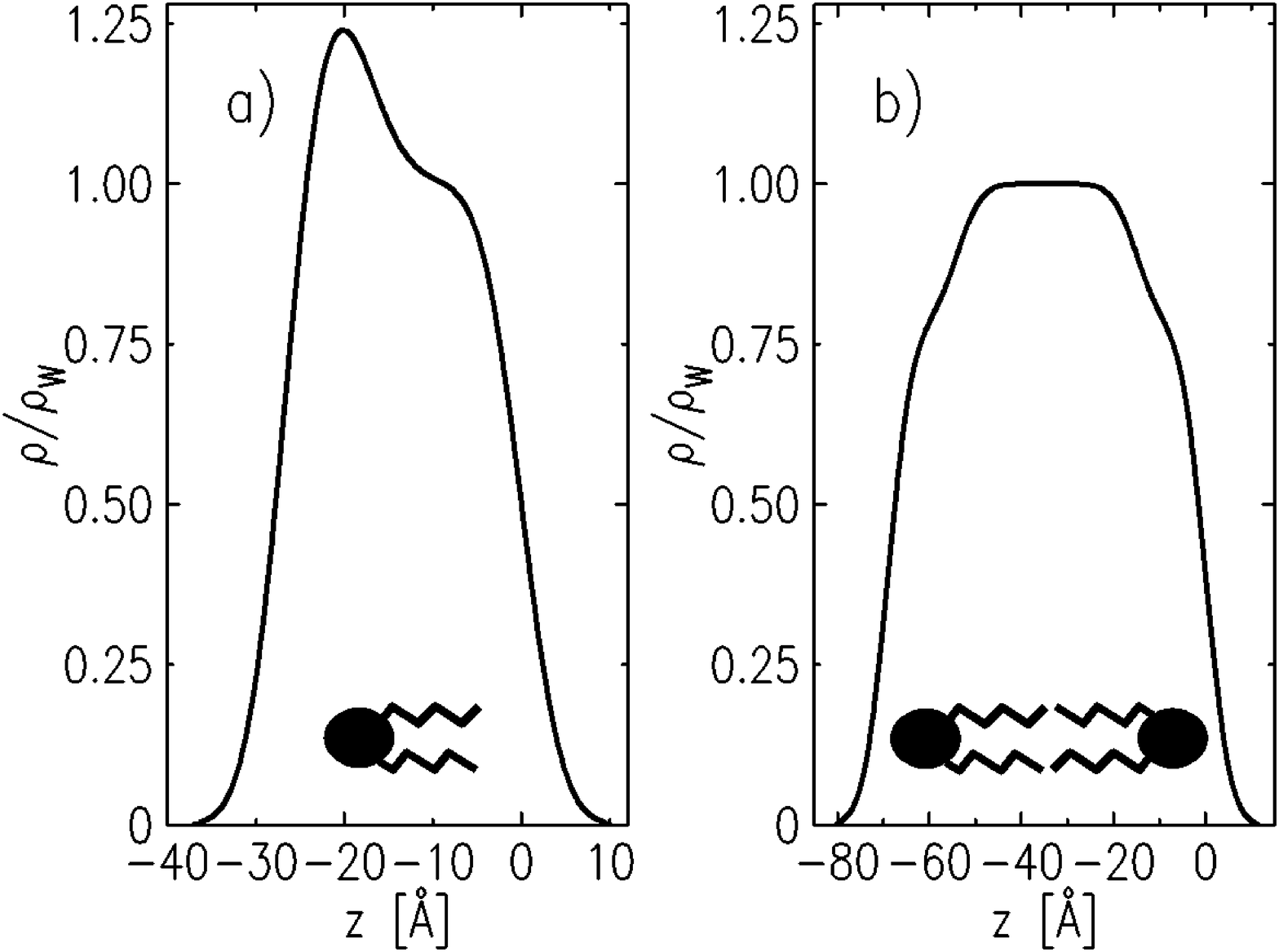, width=0.75\textwidth}

Figure 4. Model profiles $\rho(z)$ normalized to $\rho_w=0.333$ {\it e$^-$/}{\AA}$^3$
for the (a) lipid monolayer on the surface of water and
(b) the lipid bilayer in the multilayer at the hydrosol's surface.
The interfaces of the layers with air and the substrate are at $z = 0$ and $z < 0$, respectively.
\end{figure}

To conclude, the observed thickness of the adsorbed DSPC layer, $\sim 400$ \AA, corresponds to the
Debye screening length in the hydrosol bulk ($\sim 1000$ \AA)
and to the width of the transient layer on its surface,
which appears due to the difference between the potentials of the electrical image forces
for Na$^+$ cations and negatively charged nanoparticles (macroions) [20].
The spontaneous formation of the lipid multilayer is apparently attributed to the unique
boundary conditions on the surface of hydrosol on which the electric field orienting
molecular dipoles reaches $\sim 10^9 - 10^{10}$ V/m [16]. A high concentration of
Na$^+$ ions on the surface of hydrosol, $\sim 10^{19}$ m$^{–2}$, can
also promote the formation of the multilayer. The ability of sodium ions to penetrate deep into
phospholipid membranes and, thus, to form a positive surface potential,
was discussed earlier in [21, 22].

Use of the National Synchrotron Light Source, Brookhaven National Laboratory, was supported by the U.S. Department of Energy, Office of Science, Office of Basic Energy Sciences, under Contract No. DE-AC02-98CH10886.
The operation of the X19C station was supported by the ChemMatCARS Foundations
of the University of Chicago, the University of Illinois at Chicago, and Stony Brook University.
The author also thanks Grace Davison for providing Ludox solutions of colloidal silica.

\end{document}